\documentclass[journal]{IEEEtran}
\usepackage[T1]{fontenc}

\usepackage{cite}

\usepackage{multirow}
\usepackage{graphicx}
\usepackage{amssymb}
\usepackage{amsmath}
\usepackage{booktabs}
\usepackage{array}
\newcolumntype{L}[1]{>{\raggedright\let\newline\\\arraybackslash\hspace{0pt}}m{#1}}
\newcolumntype{C}[1]{>{\centering\let\newline\\\arraybackslash\hspace{0pt}}m{#1}}
\newcolumntype{R}[1]{>{\raggedleft\let\newline\\\arraybackslash\hspace{0pt}}m{#1}}

\usepackage{lipsum}
\usepackage{changepage}
\usepackage{color,soul}
\usepackage{enumitem}   

\usepackage{pifont}

\usepackage{tikz}

\usepackage{lipsum}

\let\OLDthebibliography\thebibliography
\renewcommand\thebibliography[1]{
  \OLDthebibliography{#1}
  \setlength{\parskip}{0pt}
  \setlength{\itemsep}{0pt plus 0.3ex}
}

\usepackage{array}

\makeatletter
\newcommand{\thickhline}{%
    \noalign {\ifnum 0=`}\fi \hrule height 1pt
    \futurelet \reserved@a \@xhline
}
\newcolumntype{"}{@{\hskip\tabcolsep\vrule width 1pt\hskip\tabcolsep}}
\makeatother

\usepackage{framed,multirow}

\usepackage{latexsym}

\usepackage{url}
\usepackage{xcolor}

\usepackage{hyperref}

\usepackage{bm}

\begin{document}


\title{A machine learning-based method for estimating the number and orientations of major fascicles in diffusion-weighted magnetic resonance imaging}

\author{Davood Karimi, Lana Vasung, Camilo Jaimes, Fedel Machado-Rivas, Shadab Khan, \\ Simon K. Warfield,~\IEEEmembership{Fellow,~IEEE}, and Ali Gholipour,~\IEEEmembership{Senior Member,~IEEE}

\thanks{This study was supported in part by the National Institutes of Health (NIH) grants R01 EB018988, R01 NS106030, R01 NS079788; a Technological Innovations in Neuroscience Award from the McKnight Foundation; and the Department of Radiology at Boston Children's Hospital. The content is solely the responsibility of the authors and does not necessarily represent the official views of the NIH or the McKnight Foundation. L. Vasung was supported by the Ralph Schlager Fellowship of Harvard University.

D. Karimi, C. Jaimes-Cobos, F. Machado-Rivas, S. Khan, S.K. Warfield, and A. Gholipour are with the Computational Radiology Laboratory of the Department of Radiology at Boston Children's Hospital, and Harvard Medical School, Boston, Massachusetts, USA (email: davood.karimi@childrens.harvard.edu).

L. Vasung is with the Department of Pediatrics at Boston Children's Hospital, and Harvard Medical School, Boston, Massachusetts, USA.

The code and trained models for this study are publicly available at: https://github.com/bchimagine.}
}

\maketitle

\begin{abstract}

Multi-compartment modeling of diffusion-weighted magnetic resonance imaging measurements is necessary for accurate brain connectivity analysis. Existing methods for estimating the number and orientations of fascicles in an imaging voxel either depend on non-convex optimization techniques that are sensitive to initialization and measurement noise, or are prone to predicting spurious fascicles. In this paper, we propose a machine learning-based technique that can accurately estimate the number and orientations of fascicles in a voxel. Our method can be trained with either simulated or real diffusion-weighted imaging data. Our method estimates the angle to the closest fascicle for each direction in a set of discrete directions uniformly spread on the unit sphere. This information is then processed to extract the number and orientations of fascicles in a voxel. On realistic simulated phantom data with known ground truth, our method predicts the number and orientations of crossing fascicles more accurately than several existing methods. It also leads to more accurate tractography. On real data, our method is better than or compares favorably with standard methods in terms of robustness to measurement down-sampling and also in terms of expert quality assessment of tractography results.

\end{abstract}


\IEEEpeerreviewmaketitle

\section{Introduction}

\IEEEPARstart{D}{iffusion}-weighted magnetic resonance imaging (DW-MRI) is a powerful tool for non-invasive probing of brain micro-structure, with important applications in studying brain connectivity, development, and degeneration \cite{johansen2013}. In brain connectivity studies, DW-MRI measurements are used to infer the number and orientations of major fascicles or a fiber orientation distribution function (fODF) in each voxel. Tractography techniques are then used to create a brain connectivity map from these local information. Early studies relied on the diffusion tensor imaging (DTI) model, which modeled the anisotropic diffusion in each voxel with a single tensor \cite{basser1994}. Limitations of this model were evident from the outset, as it failed to accommodate voxels with multiple fascicles. Such voxels account for a large fraction of brain white matter \cite{alexander2001}. Accurate tractography and connectivity analysis require more flexible models that can estimate the number and orientations of multiple fascicles in a voxel.

Existing methods for estimating the number and orientations of fascicles in multi-fascicle voxels can be divided into parametric and non-parametric methods \cite{seunarine2014}. Parametric methods model the diffusion signal as the sum of signals from different compartments and estimate the model parameters using a model fitting method. This is usually performed with the use of nonlinear optimization techniques. Non-parametric methods, on the other hand, typically estimate the probability distribution of diffusion or of fiber orientations on a sphere. Prominent peaks of the estimated distribution are assumed to correspond to major fascicles. However, both techniques have important shortcomings \cite{seunarine2014}. Parametric methods suffer from local minima and sensitivity to initialization as they involve solving a non-convex optimization problem. Furthermore, determining the correct number of fascicles in each voxel is not straight-forward and depends on additional methods, which are often complicated and computationally expensive \cite{schultz2010,scherrer2013}. For non-parametric models, the choice of the right representation for the distribution on sphere is unclear. Most methods use linear representations, which are not sufficiently complex \cite{seunarine2014}. Moreover, they often produce spurious peaks that can be hard to distinguish from the true ones \cite{tournier2008}.

A less common but promising class of methods are data-driven and machine learning-based techniques. With the growing diversity and power of machine learning models and techniques, these methods have been gaining more attention in recent years. Several recent studies have attempted at estimating scalar diffusion parameters such as diffusion kurtosis measures and generalized fractional anisotropy using deep learning \cite{golkov2016,gibbons2019simultaneous,ye2019deep,aliotta2019highly}. They have shown that deep learning methods can accurately estimate such parameters from highly under-sampled q-space data. However, the above-mentioned studies do not address estimation of the number and orientations of fascicles, which is the aim of this paper.

A number of studies have proposed machine learning methods for estimating the number and/or orientations of major fascicles or for estimating the complete fODF. Support vector regression (SVR) was used to estimate the number of fascicles in each voxel in \cite{schultz2012}. Another study used convolutional neural networks (CNNs) for the same purpose \cite{koppers2017reliable}. Another recent study proposed using CNNs for estimating the orientations of fascicles \cite{koppers2016}. However, they stacked one-dimensional diffusion measurements in an artificial manner to synthesize 2D input signals for their CNN. Other CNN-based methods work on small patches of DW-MRI images as input \cite{lin2019fast,koppers2017reconstruction}. In a different method, fODF was estimated in a sparse signal reconstruction framework, wherein a deep learning model was used to estimate the sparse reconstruction coefficients \cite{ye2017fiber}. One study learned an fODF prior using auto-encoders and incorporated this prior within more traditional optimization-based fODF estimation techniques \cite{patel2018better}. Some studies represent the diffusion signal and/or the fODF in spherical harmonic bases and use machine learning methods to estimate the coefficients of the fODF from those of the diffusion signal \cite{nath2019,nath2019deep}. The latter methods used ground truth training data with histological tracing of fiber orientations, which is very hard to come by. 

Machine learning methods have also been used for tractography \cite{poulin2019tractography}. Many of these methods learn to perform tractography directly based on the raw diffusion data without first explicitly estimating an fODF or orientations of major fascicles. For example, random forest classifiers were trained to estimate the next step in a streamline tractography process \cite{neher2015machine,neher2017fiber}. This method was shown to produce results that were comparable with or better than standard tractography techniques. Other studies have successfully used deep learning models such as recurrent neural networks \cite{benou2019deeptract,jorgens2018learning}.

In this paper, we propose a novel method for estimating the number and orientations of fascicles in each voxel from DW-MRI measurements. Although our method makes use of a deep learning model, it is quite different from all prior studies. Unlike parametric methods that aim to optimize all model parameters jointly, and unlike non-parametric methods that estimate the orientation distribution function on the entire sphere at once, our method uses all measurements in a voxel to estimate one single parameter at a time. Specifically, as we explain in detail below, we consider a set of directions on the unit sphere. For each direction in this set, one at a time, we use all DW-MRI measurements in the voxel to estimate the angle to the closest fascicle for that direction. This information can be further processed to estimate the number and orientations of major fascicles or used directly for tractography. Our method can be trained using either simulated or real DW-MRI data. We show that our method achieves results that are comparable with or better than several competing methods.

\section{Materials and methods}
\label{methods}

\begin{figure*}[htb]

  \centering
 \includegraphics[width=0.8\textwidth]{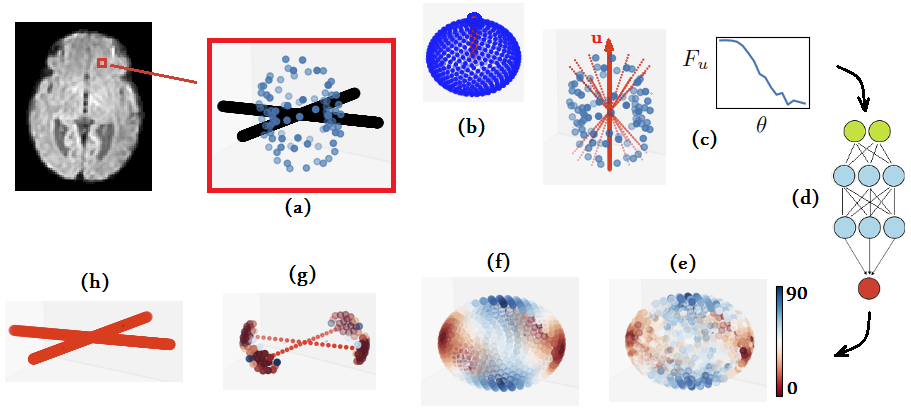}

\caption{A schematic summary of our proposed method for estimating the number and orientations of major fascicles. Given the diffusion measurements in an imaging voxel \textbf{(a)}, we consider a set of directions on the unit sphere \textbf{(b)}. For each direction on the unit sphere, $u$, we compute \textbf{(c)} a feature vector $F_u(\theta)$ using Equation \eqref{eq:feature_vector}. We use a multi-layer perceptron, \textbf{(d)}, to estimate the angle to the closest fascicle, $\Phi(u)$, based on this feature vector. This is performed separately for each direction on the unit sphere, \textbf{(e)}. The estimated angles, $\Phi(u)$, are smoothed using bivariate spline smoothing to obtain a more accurate estimation of the angle to the closest fascicle $\Phi_s(u)$, \textbf{(f)}. We then threshold $\Phi_s(u)$ and perform local minimum extraction to determine the number and approximate orientations of the fascicles, \textbf{(g)}. Finally, the estimated orientations are further refined using Karcher mean to obtain a more accurate estimation of fascicle orientations, \textbf{(h)}.}
\label{fig:method_schematics}
\end{figure*}

Figure \ref{fig:method_schematics} shows an overview of our method. In brief, for each voxel, we consider a set of directions on the unit sphere. For each direction, we estimate the angle to the closest fascicle. We then process this information to determine the number and orientations of major fascicles in that voxel. We explain the steps of our method below. To simplify the presentation of our proposed method, we assume a single-shell measurement scheme, i.e., only one gradient strength, $b$.

\subsection{Feature vector computation}

Let us denote the set of $m$ DW-MRI measurements in a voxel with $\{ s_i(q_i, b) \}_{i=1}^m$, where the unit vector $q_i$ is the gradient direction for the $i^{\text{th}}$ measurement. Given an arbitrary direction $u$ on the unit sphere, our intermediate goal is to estimate the angle of the closest fascicle to $u$.

We begin by noting that, assuming axially-symmetric fascicles, the diffusion signal can be modeled as \cite{anderson2005}:
\begin{equation} \label{eq:signal_model} 
\begin{aligned}
s_i(q_i)= s_0 \Bigg( & f_{\textrm{iso}} \exp(-b \lambda_{\textrm{iso}}) +\\
& \sum_{k=1}^K f_k \exp \bigg( -b \big(\lambda_{\perp}^k \\
& \hspace{20mm}+ 3 (\bar{\lambda}^k - \lambda_{\perp}^k ) \cos^2 \alpha_i^k \big)  \bigg)   \Bigg)
\end{aligned}
\end{equation}
\noindent where $s_0$ is the signal when no diffusion gradient is applied. Also, $\bar{\lambda}^k= ( \lambda_{\parallel}^k + 2 \lambda_{\perp}^k )/3$, where $\lambda_{\parallel}^k$ and $\lambda_{\perp}^k$ denote the axial and radial diffusivities, respectively, for the $k^{\text{th}}$ fascicle, and $\lambda_{\textrm{iso}}$ is the diffusivity of the isotropic compartment. Moreover, $\alpha_i^k$ is the angle between the $k^{\text{th}}$ fascicle and $q_i$. Finally, $K$ is the number of fascicles crossing the voxel and $f_k$s denote the occupancy fraction of each compartment. From the above equation, because all parameters except for $\alpha$'s are fixed in a voxel, we observe that the signal in any direction is mainly a function of the angles between that direction and the fascicles, with the closest fascicle having the largest influence.

\begin{figure*}[htb]
  \centering
\includegraphics[width=0.85\textwidth]{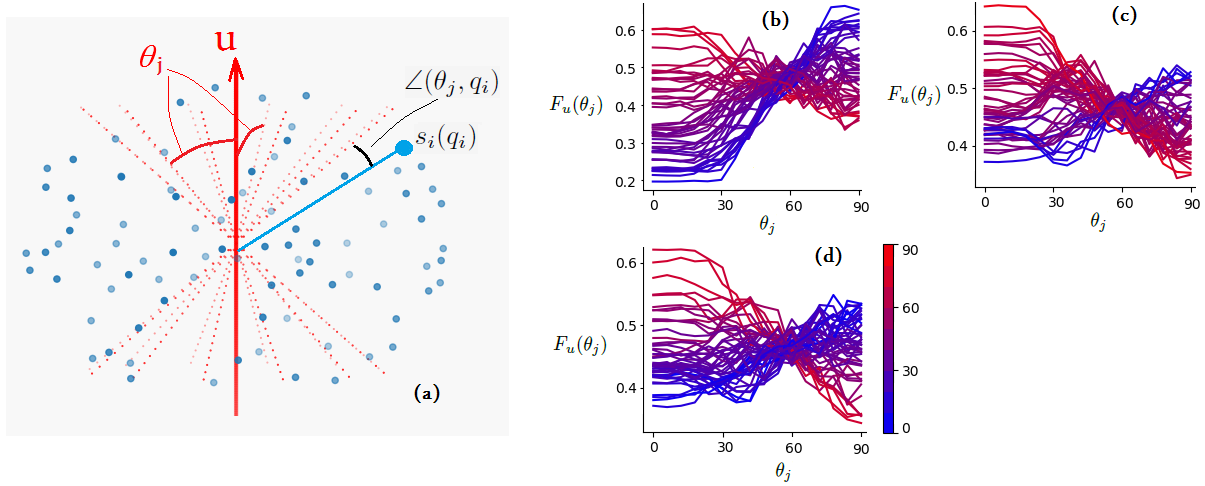}
\caption{ \textbf{(a)} This schematic shows how Equation \eqref{eq:feature_vector} computes the feature vector for a direction $u$. For simplicity of illustration, the direction $u$ is chosen as the vertical direction in this figure. Geometrically speaking, for each angle $\theta_j$ we consider a cone with that angle around $u$. An example cone is shown with dotted red lines. We compute the weighted average of the diffusion signal, $s(q_i)$, with weights depending on the closeness of $q_i$ to the cone. The right side of the figure displays feature vectors for 50 randomly selected directions for voxels with 1-3 fascicles (\textbf{(b)}-\textbf{(d)}, respectively). The colorbar shows the angle of $u$ to the closest fascicle. In other words, the blue curves are feature vectors for directions $u$ that are very close to a major fascicle, whereas the red curves are for directions that make an angle of almost $90^\circ$ with the closest fascicle. It is clear that the proposed feature vector, $F_u(\theta_j)$, is strongly related to the angle of $u$ to its closest fascicle.}
\label{fig:fvector_samples}
\end{figure*}

Based on the above argument, we propose the following feature vector to be used for estimating the angle to the closest fascicle for an arbitrary direction $u$:

\begin{equation} \label{eq:feature_vector} 
\begin{aligned}
&F_u(\theta_j)= \sum_i \omega( \angle( \theta_j, q_i)) s_i (q_i)/s_0  \\
& \hspace{10mm} \theta_j= j\pi/(2n), \hspace{3mm} j=0:n
\end{aligned}
\end{equation}

\noindent where $\theta$ is the angle away from $u$, and $\angle(\theta_j, q_i)$ is the angle between $\theta_j$ and $q_i$. This is simply a weighted average of the diffusion measurements as a function of $\theta$. Please see Figure \ref{fig:fvector_samples}(a) for a schematic illustration. Simply put, for each angle $\theta_j$ we consider a cone with that angle around $u$, and compute the weighted average of the diffusion signal measurements $s(q_i)$, with weights depending on the closeness of $q_i$ to the cone. We used $\omega \propto 1/ ( \angle( \theta_j, q_i) + \epsilon )$ with $\epsilon=0.1$, in order to give larger weights to measurements with $q$ closer to $\theta_j$. We compute this feature vector for a set of $n+1$ angles $\theta_j$ from 0 to $\pi/2$. We set $n=15$ in this work. Figure \ref{fig:fvector_samples}(b)-(d) shows example feature vectors for arbitrary directions in voxels with 1, 2, and 3 fascicles.

\subsection{Estimation of the angle to the closest fascicle}

As discussed above and illustrated in Figure \ref{fig:fvector_samples}, the computed feature vector is strongly related to the angle to the closest fascicle. Therefore, we propose to estimate the angle to the closest fascicle based on the computed feature vector. We used a multi-layer perceptron (MLP) for this purpose because MLPs have a high representational capacity and our experiments showed that they produced much more accurate predictions than competing models such as SVR. Based on preliminary cross-validation experiments, we decided on a network with six hidden layers, with $\{ 30, 60, 80, 80, 60, 30 \}$ neurons in the hidden layers. The input layer is of size equal to the feature vector length, i.e., $n+1$, and the output is a scalar, representing the angle to the closest fascicle.

\subsection{Predicting the number and orientations of fascicles}
\label{postprocessing}

We use the trained MLP to estimate the angle to the closest fascicle for a set of directions uniformly spread on the unit sphere. In this work we used 724 directions, resulting in a resolution of approximately $7^{\circ}$. We denote these predictions with $\Phi(u)$. We process $\Phi(u)$ in three steps to determine the number and orientations of fascicles in a voxel. These steps are illustrated in Figure \ref{fig:method_schematics}(e)-(h) and they are explained below.

\begin{enumerate}
    
    \item \textbf{Smoothing to reduce the estimation error in $\Phi(u)$.} The error in the estimated angles, $\Phi(u)$, can be reduced by exploiting the knowledge that the estimated angles for nearby directions should be close. For this step, we use bivariate spline smoothing in spherical coordinates \cite{dierckx1995}. This operation results in a smoother and more accurate estimate of the angle to the closest fascicle, which we denote with $\Phi_s(u)$.
    
    \item \textbf{Thresholding to identify candidate fascicle orientations.} Directions $u$ for which the estimated angle to the closest fascicle, $\Phi_s(u)$, is close to zero are candidates for fascicle orientations. Hence, we define $u_t= \{u | \Phi_s(u)<t \}$ as the candidates. Based on the resolution of our spherical grid ($7^{\circ}$) and the accuracy of $\Phi_s$ (discussed below), we chose a threshold of $t= 30^{\circ}$. However, this procedure generates a large number of candidate fascicle orientations, many of which will be clustered together around the true fascicle orientations. Therefore, within the set $u_t$ we perform a local-minimum extraction to detect the true fascicles. We mark a direction $u^* \in u_t$ as a local minimum if $\Phi_s(u^*)$ is not larger than any of its neighbors and smaller than at least one of its neighbors. We used the DIPY package \cite{garyfallidis2014} for local minimum extraction.
    
    \item \textbf{Estimating fascicle orientations.} From the above step we obtain a set of estimated fascicle orientations, $\{ u^* \}$. The number of vectors in this set is our estimate of the number of major fascicles in the voxel. One can propose the directions of $\{ u^* \}$ as the estimated fascicle orientations. However, we can improve the estimated fascicle orientations by exploiting not only the directions of local minima, but also the nearby directions in $u_t$. Therefore, we further improve the estimated fascicle orientations as follows. For each $u^*_i \in \{ u^* \}$, we identify all candidate directions in $u_t$ that are closer to $u^*_i$ than to any other direction in $\{ u^* \}$. Let us denote the set of these directions with $\{ u_t^i \}$. We compute the Karcher mean \cite{pennec1999} of the directions in the set $\{ u_t^i \}$ to obtain a more accurate estimate of the orientation of the fascicle $u^*_i$.

\end{enumerate}

\subsection{Implementation details and compared methods}

We implemented our method in Python and TensorFlow, and compared it with five other methods explained below.

\begin{itemize}

\item Bayesian method of \cite{behrens2007,behrens2003characterization}. This method uses automatic relevance determination (ARD) to estimate the number of major fascicles. All model parameters including the orientation of major fascicles are estimated in a Bayesian framework using Metropolis Hastings Markov Chain Monte Carlo sampling.

\item Multi-tensor model fitting with F-test for model selection. F-test is a well-known statistical model selection method that has been used by previous studies in DW-MRI modeling \cite{alexander2002detection,kreher2005multitensor,scherrer2012parametric}. In this work, we used it for selecting the number of fascicles in a multi-tensor model with an additional isotropic compartment, similar to Equation \eqref{eq:signal_model}.

\item Multi-tensor model fitting with selection of the number of fascicles based on generalization error. We used the method proposed in \cite{scherrer2013}. This method is based on representing each of the fascicles with a tensor. The number of fascicles is determined by estimating the generalization error using the 632+ bootstrap technique \cite{efron1997improvements}.

\item Constrained spherical deconvolution (CSD) \cite{tournier2007}. This method estimates the fODF on a spherical grid. To be consistent, we used the same sphere as that used for our proposed method, which included 724 directions. In order to estimate the number of major fascicles from the estimated fODF, we used the method proposed in \cite{schultz2010}. This method depends on a set of thresholds to decide whether two or more fascicles exist in a voxel. We determined these thresholds using cross-validation on a digital phantom with known ground-truth.

\item Sparse Fascicle Model (SFM) \cite{rokem2015}. Similar to CSD described above, for SFM we used the method of \cite{schultz2010} to determine the number of major fascicles.

\end{itemize}

Similar to the method of \cite{schultz2010} used with CSD and SFM, both of the tensor fitting methods also involved thresholds for estimating the number of fascicles \cite{scherrer2013}. We optimized the thresholds using data simulated on a phantom similar to that in \cite{scherrer2013}. We used the same phantom data to select the hyper-parameters of the Monte Carlo method used for the ARD method, \cite{behrens2007}, and to tune our proposed method, which mainly included setting the value of the threshold $t$.

We used either simulated or real diffusion data to train our model, as explained below. To generate simulated diffusion data, we used a multi-tensor model according to Equation \eqref{eq:signal_model} with $K$ up to 3 and range of diffusivity values $\lambda_{\parallel} \in [0.0018,0.0024]$ and $\lambda_{\perp} \in [0.00035, 0.00050]$. For training with real diffusion data, we used 75 scans from the developing Human Connectome Project (dHCP) dataset \cite{bastiani2019}. Diffusion scans in this dataset include 300 measurements with diffusion strength, $b$, of 10 (n=20), 400 (n=64), 1000 (n=88), and 2600 (n=128). To generate a pseudo ground truth for the number and orientations of major fascicles on the 75 training scans, we applied CSD on all 300 diffusion measurements. We then applied our trained model on a separate set of 20 test scans from the same dataset. On the test scans, we only used the 88 measurements in the $b=1000$ shell.

\subsection{Tractography and semi-quantitative analysis of fiber tracts}
\label{tractography_eval_description}

In some of our experiments with phantom and real data, we compared different methods in terms of tractography. In those experiments, we used a standard tractography algorithm from DIPY \cite{garyfallidis2014} with the same (default) tractography settings for all methods.

In our experiments with real DW-MRI data, an expert neuroanatomist (LV) compared our method with other competing methods in terms of the quality of generated tractograms. This evaluation was based on assessment of commissural (corpus callosum and anterior commissure), projection (frontopontine fibers, corticospinal tract, and fornix), association (cingulum, inferior fronto-occipital fasciculus, and uncinate fasciculus), and cerebellar (middle cerebellar pedunculus) tracts. The tracts were first visualized using the default track group created with three coronal (Y) slice filters. The filters were situated in the frontal, central, and parietal regions of the brain. The frontal coronal slice filter was situated in the white matter rostral and adjacent to the most anterior tip of the lateral ventricles and was used to visualize cingulum, corpus callosum, frontopontine fibers, uncinate fasciculi, and inferior fronto-occipital fasciculi. The central slice filter was situated at the level of midline bundle component of the anterior commissure and was used to visualize the corticospinal tract, anterior commissure, and fornices.  The parietal slice filter was situated in the white matter caudal and adjacent to the posterior tip of the splenium of the corpus callosum and was used to visualize middle cerebellar pedunculi and cingulum.

Next, the white matter tracts were visually inspected and graded based on their integrity (i.e., all components of the tract being visible) and bilateral presence. Grade 3 was characterized by the presence of all components of the given tract bilaterally. Grade 2 was characterized by the absence of at least one component of the given tract regardless of the side. Grade 1 was characterized by the complete absence of the tract in both hemispheres of the brain. One exception to this grading rule was the corticospinal tract. For the corticospinal tract, we assessed only its caudal part, coursing between cerebral peduncles and decussation of pyramids, which we treated as a separate tract.

\section{Results and Discussion}
\label{results}

\subsection{Preliminary cross-validation experiments}

We performed the preliminary validation of our proposed method on a digital phantom similar to that in \cite{scherrer2013}. This is a $15 \times 15$-voxel phantom with 1, 2, or 3 fascicles per voxel. In these experiments, the root mean square of the error in the angle to the closest fascicle estimated by our MLP was 7.0, 7.4, and 8.2 degrees for voxels with 1, 2, and 3 fascicles, respectively. After refining the estimations with the steps explained in Section \ref{postprocessing}, these errors decreased to 3.2, 3.7, and 5.7 degrees, respectively. For detecting voxels with 1, 2, and 3 voxels, our method achieved accuracy of 0.98, 0.98, and 0.90, respectively.

\subsection{Evaluation on the HARDI-2013 phantoms}
\label{hardi_evaluation}

One of the phantoms in this dataset contains voxels with up to five crossing fascicles \cite{caruyer2014}. We used the HARDI scheme of this data, which included 64 gradient directions at b = 3000. We do not know the model and range of diffusivities used to synthesize this data. To train our MLP, we simulated data using the multi-tensor model (Equation \eqref{eq:signal_model}). We used the parameter settings optimized in the training step above for all methods including ours.

Table \ref{table:MoSe} shows the comparison of different methods in terms of accuracy, sensitivity, and specificity in detecting voxels with one, two, and three fascicles. For this evaluation, we considered voxels for which the angle between neighboring fascicles was as low as $30 ^{\circ}$ and fraction of each fascicle was as low as 0.15. Our method achieved higher accuracy values than all other methods in detecting voxels with one, two, and three voxels. In terms of sensitivity and specificity, our method achieved results that were better than or competitive with other methods. In detecting voxels with 2 or 3 fascicles our method performed much better than all other methods. Most other methods completely failed in detecting voxels with three fascicles, whereas our method was accurate.

\begin{table*}[!htb]
\footnotesize
  \begin{center}
    \begin{tabular}{L{2.8cm} C{1.1cm} C{1.2cm} C{1.3cm} C{1.1cm} C{1.2cm} C{1.3cm} C{1.1cm} C{1.2cm} C{1.3cm} }
\hline
 &  \multicolumn{3}{c}{One fascicle} &  \multicolumn{3}{c}{Two fascicles} &  \multicolumn{3}{c}{Three fascicles} \\ 
 Method & Accuracy & Sensitivity & Specificity & Accuracy & Sensitivity & Specificity & Accuracy & Sensitivity & Specificity \\ \hline
SFM        & $0.62$ & $\textbf{1.00}$ & $0.32$ & $0.74$ & $0.42$ & $\textbf{1.00}$ & $0.87$ & $0.00$ & $0.00$ \\
CSD        & $0.90$ & $\textbf{1.00}$ & $0.85$ & $0.81$ & $0.83$ & $0.67$ & $0.82$ & $0.43$ & $1.00$ \\
Tensor fitting- F-test     & $0.93$ & $0.94$ & $0.93$ & $0.68$ & $0.96$ & $0.51$ & $0.67$ & $0.00$ & $0.00$ \\

Tensor fitting- bootstrap       & $0.88$ & $0.66$ & $\textbf{0.99}$ & $0.55$ & $0.98$ & $0.43$ & $0.67$ & $0.00$ & $0.00$ \\
ARD     & $0.92$ & $0.87$ & $0.94$ & $0.74$ & $0.90$ & $0.58$ & $0.78$ & $0.38$ & $0.87$ \\
Proposed method           & $\textbf{0.97}$ & $0.99$ & $0.96$ & $\textbf{0.93}$ & $\textbf{0.98}$ & $0.85$ & $\textbf{0.92}$ & $\textbf{0.75}$ & $\textbf{0.98}$ \\
\hline 
    \end{tabular}
  \end{center}
  \caption{\footnotesize{Accuracy, sensitivity, and specificity of different methods in detecting voxels with 1-3 fascicles. The best results in each column have been marked using bold type. Overall, our method achieved the best performance.}}
  \label{table:MoSe}
\end{table*}

We also compared the methods in terms of their accuracy in estimating fascicle orientations by computing the weighted average angular error (WAAE) \cite{schultz2012}, defined for each voxel as $\text{WAAE}= \sum_i w_i \underset{j}{\min} \arccos(|v_i, \hat{v}_j |)$, where $v$ and $\hat{v}$ denote the true and estimated fascicle orientation direction vectors and $w_i$ is the true occupancy fraction of the $i^{\text{th}}$ fascicle. In other words, WAAE weights the errors in estimating fascicle orientations in a way that more prominent fascicles are assigned larger weights. Table~\ref{table:WAAE} reports the values of WAAE separately for all voxels containing one, two, and three fascicles in this phantom. For voxels with two or three fascicles, our method achieved more accurate estimation of fascicle orientations than all other methods. For voxels with one fascicle, our method was less accurate than some of the methods such as CSD and SFM. However, it is likely that fascicle orientation estimation accuracy in voxels with multiple fascicles is more important for the purposes of tractography and connectivity analysis. It is also noteworthy that a finer spherical grid, i.e., a larger number of directions $u$ on the unit sphere, would increase the spatial resolution, which may further improve the accuracy of our method in estimating fascicle orientations.

\begin{table}[!htb]
\footnotesize
  \begin{center}
    \begin{tabular}{L{3.0cm} C{1.3cm} C{1.5cm} C{1.3cm}  }
\hline
Method &  One fascicle &  Two fascicles &  Three fascicles \\  \hline
SFM                       & $2.99$ & $11.7$ & $17.8$ \\
CSD                       & $\textbf{2.85}$ & $11.8$ & $12.8$ \\
Tensor fitting- F-test    & $5.76$ & $10.6$ & $15.4$ \\
Tensor fitting- bootstrap & $3.00$ & $10.2$ & $15.1$ \\
ARD                       & $4.58$ & $9.31$ & $12.8$ \\
Proposed method           & $4.05$ & $\textbf{9.23}$ & $\textbf{10.9}$ \\
\hline 
    \end{tabular}
  \end{center}
  \caption{\footnotesize{Comparison of different methods in estimating fascicle orientations in terms of WAAE, in degrees, in voxels with 1-3 fascicles. The best results in each column have been marked using bold type.}}
  \label{table:WAAE}
\end{table}

For another phantom in this dataset, 20 pairs of seed and target tractography regions of interest (ROIs) are known. On this phantom, we applied different methods to estimate the number and orientations of major fascicles or the fODF in each voxel, followed by tractography. For tractography, we launched one streamline from each seed voxel and computed the fraction of the streamlines that ended within a distance of at most two voxels from a corresponding target voxel. We refer to this fraction as success ratio. Table \ref{table:success_ratio} compares different methods in terms of success ratio. Our method achieved a higher average success ratio than other methods. SFM and CSD also achieved high success ratios in this experiment.

\begin{table}[!htb]
\footnotesize
  \begin{center}
    \begin{tabular}{L{3.0cm} C{4.5cm}  }
\hline
Method &  success ratio $ \hspace{10mm}  \text{mean} \pm \text{std},  \hspace{8mm} [\text{min}, \text{max}]$ \\  \hline
SFM                       & $0.59 \pm 0.26, \hspace{6mm}  [0.00, \bm{0.96}]$  \\
CSD                       & $0.62 \pm 0.20, \hspace{6mm}  [0.09, 0.93]$  \\
Tensor fitting- F-test    & $0.47 \pm 0.22, \hspace{6mm}  [0.09, 0.90]$  \\
Tensor fitting- bootstrap & $0.49 \pm 0.23, \hspace{6mm}  [0.05, 0.88]$  \\
ARD                       & $0.56 \pm 0.22, \hspace{6mm}  [0.06, 0.88]$  \\
Proposed method           & $\bm{0.63 \pm 0.21}, \hspace{4mm}  [\bm{0.12}, 0.93]$  \\
\hline 
    \end{tabular}
  \end{center}
  \caption{\footnotesize{Mean, standard deviation and range of success ratio for HARDI-2013 phantom tractography.}}
  \label{table:success_ratio}
\end{table}

\subsection{Evaluation of estimated fODF with real DW-MRI data}

\begin{figure}[htb]
\begin{minipage}[b]{1.0\linewidth}
  \centering
  \centerline{\includegraphics[width=1.0\textwidth]{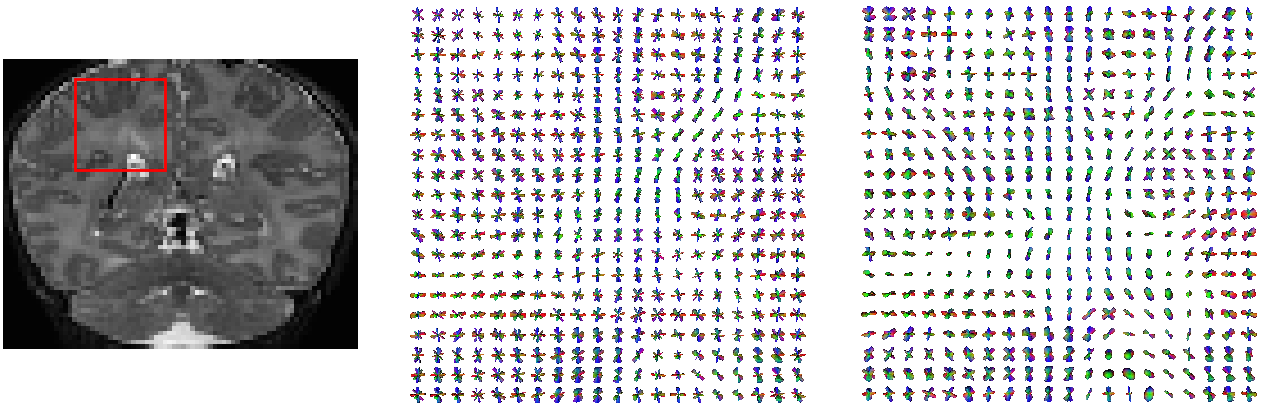}}
\end{minipage}
\caption{An example of the fODF reconstructed by our method and CSD. Left: the corresponding slice from the T2-weighted image with the ROI marked in red. The middle and right, the fODF reconstructed in that ROI by CSD and by our method, respectively.}
\label{fig:fODF}
\end{figure}

We applied our method and CSD on 20 subjects from the dHCP data \cite{bastiani2019}. For each subject, we used the 88 diffusion measurements with $b= 1000$. In the absence of ground truth, first we compared our method and CSD visually in terms of the reconstructed fODF. For our method, we propose $\text{fODF}(u)= 1/\Phi_s(u)^p$, where we set $p=2$ in the results shown below. The justification for this definition is simple; directions $u$ that are closer to major fascicles should have a higher fODF value. Increasing $p$ makes the fODF pointier without changing the directions of the peaks. A representative example of reconstructed fODFs in an area of crossing fibers has been shown in a coronal slice of a dHCP subject in Figure \ref{fig:fODF}.

For a quantitative comparison, we evaluated the robustness of the two methods to measurement down-sampling. For each subject, we removed 25\% and 50\% of the measurements at random, leaving 66 and 44 measurements, respectively. We applied our method and CSD on these down-sampled measurements and determined the orientation of the most dominant fascicle (i.e., the largest peak of the fODF). We computed the angular error between this orientation and the orientation estimated using all 88 measurements for all voxels in the white matter. This included a total of more than 1.4 million voxels. Results of this experiment are presented in Table \ref{table:downsampling}, showing that our method is more robust to measurement down-sampling than CSD. Paired t-tests at $p=0.01$ significance level showed that the angular error for our method was significantly smaller than that of CSD at both 25\% and 50\% down-sampling rates.

\begin{table*}[!htb]
\footnotesize
  \begin{center}
    \begin{tabular}{L{3.6cm} C{1.8cm} C{1.8cm} C{1.8cm} C{1.8cm}  }
\hline
 &  \multicolumn{2}{c}{Down-sampling by 25\%} &  \multicolumn{2}{c}{Down-sampling by 50\%} \\ 
&  mean $\pm$ std. & maximum & mean $\pm$ std. & maximum  \\ \hline
CSD                & $11.8 \pm 3.1$  & 22.0 & $16.6 \pm 3.2$ & 22.2 \\
Proposed method    & $\mathbf{10.9 \pm 0.7}$  & \textbf{12.6} & $\mathbf{15.8 \pm 0.5}$ & \textbf{16.9}  \\
\hline 
    \end{tabular}
  \end{center}
  \caption{\footnotesize{Comparison of the proposed method and CSD in terms of error in the estimated orientation of the most prominent fascicle in 20 subjects from the dHCP dataset at two different measurement down-sampling rates.}}
  \label{table:downsampling}
\end{table*}

\subsection{Evaluation of tractography with real DW-MRI data}

\begin{figure*}[htb]
  \centering
\includegraphics[width=1.0\textwidth]{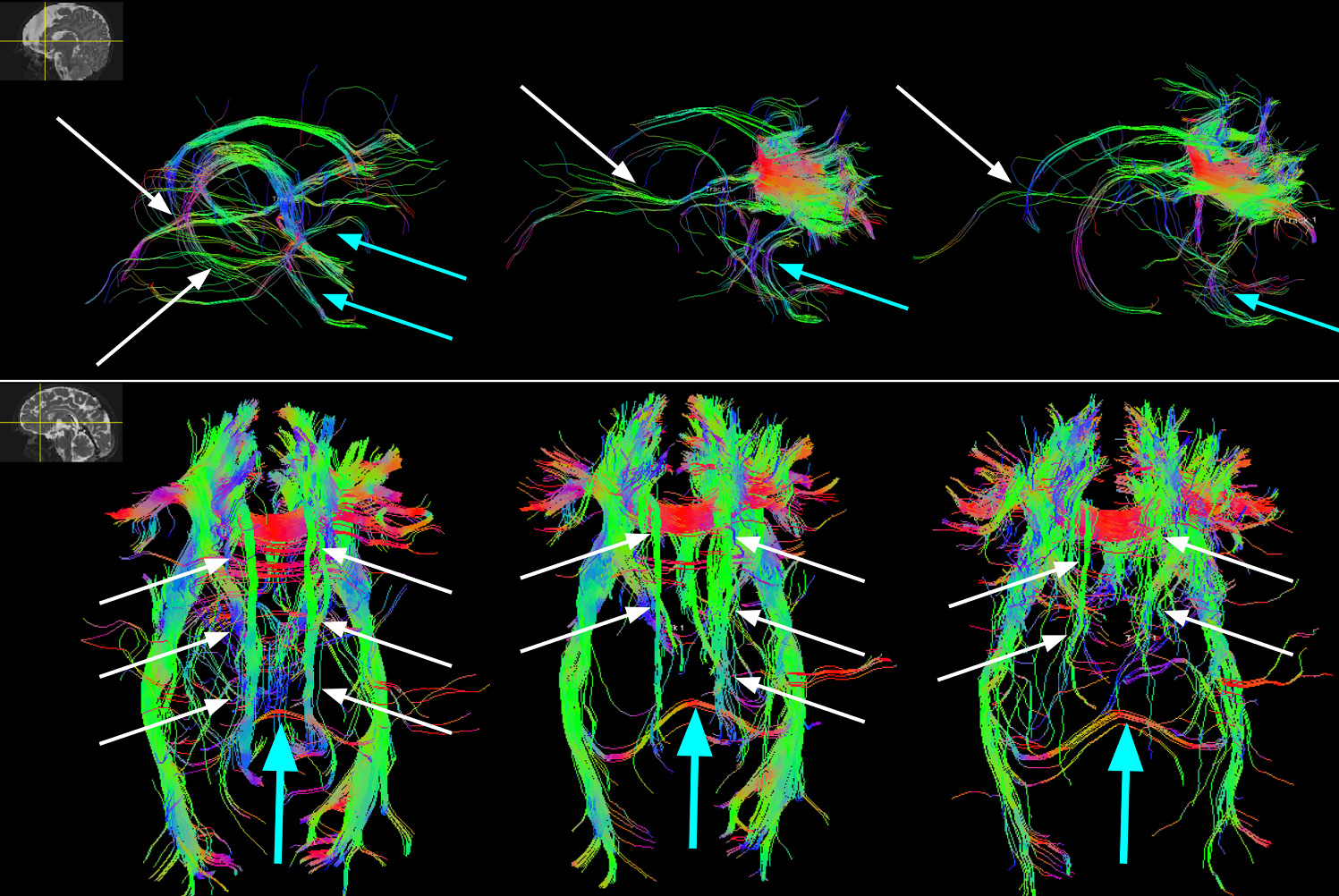}
\caption{Visualization and grading of white matter tracts for our method (left column), SFM (middle column), and DTI (right column) using the frontal slice filter (vertical yellow line in the left upper corner of each row).
Upper row: Note the integrity and bilateral presence of inferior fronto-occipital fasciculus (white arrows) and uncinate fasciculus (blue arrows) when using our method (left column). In contrast, notice the unilateral absence of inferior fronto-occipital fasciculus or uncinate fascicle when using SFM (middle column) or DTI method (right column).
Bottom row: Note the integrity and bilateral presence of cingulum (white arrows) and the absence of spurious fibers in the splenium of corpus callosum (blue arrow) when using our method (left column). In contrast, notice the unilateral or bilateral absence of components of cingulum and presence of spurious fibers in the splenium of corpus callosum when using SFM (middle column) or DTI method (right column).}
\label{fig:tractography_1}
\end{figure*}

\begin{figure*}[htb]
  \centering
\includegraphics[width=1.0\textwidth]{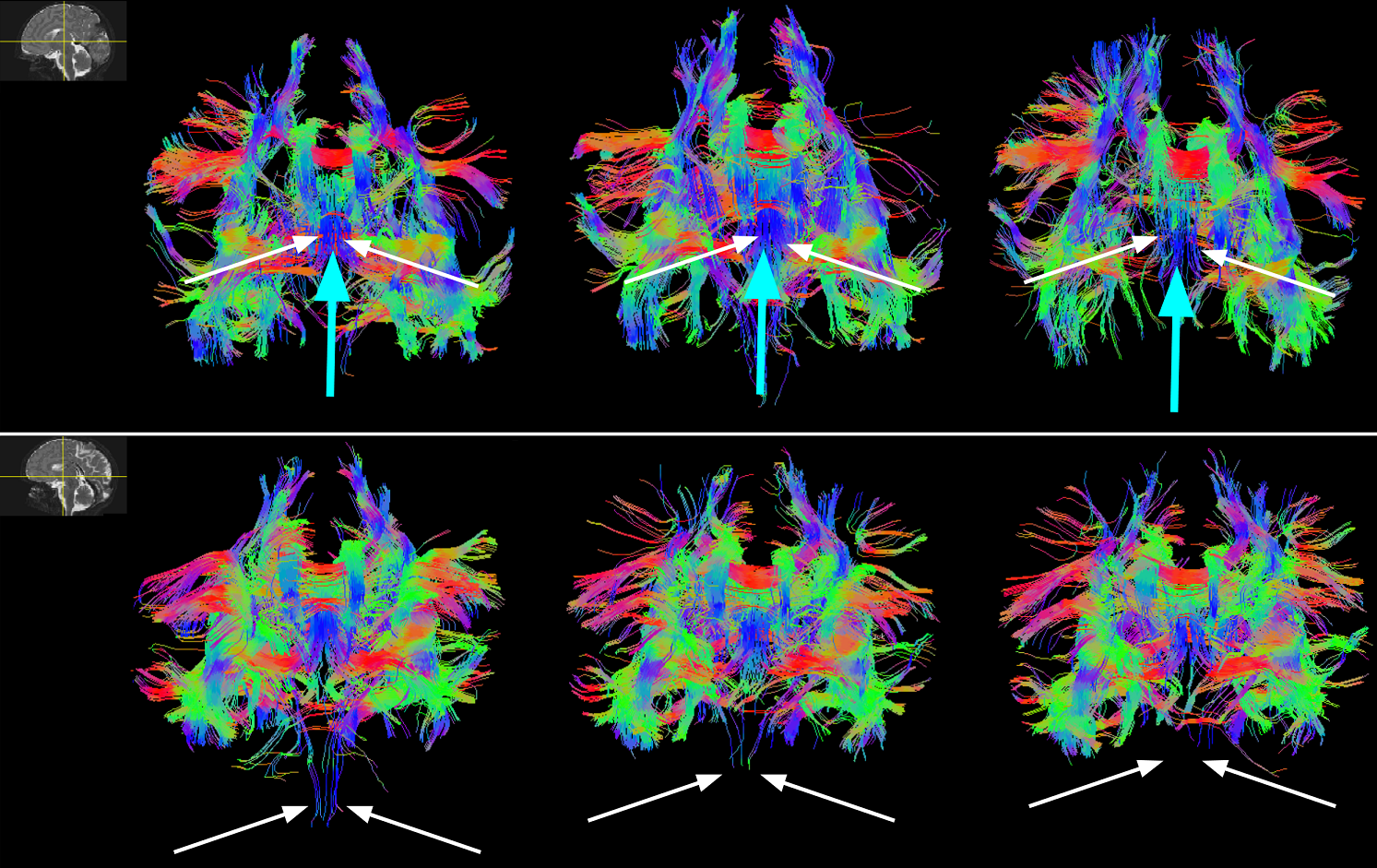}
\caption{Visualization and grading of white matter tracts for our method (left column), SFM (middle column), and DTI (right column) using the central slice filter (vertical yellow line in the left upper corner of each row). Upper row: Note the integrity and bilateral presence of all parts of the fornix (in particular columns of fornices, white arrows) and anterior commissure (middle bundle component shown with blue arrow) when using our method (left column). In contrast, notice the unilateral absence of the middle bundle component of the anterior commissure (blue arrow) when using SFM (middle column) or DTI method (right column). Bottom row: Note the integrity and bilateral presence of the caudal part of the corticospinal tract (white arrows) that reached the decussation of pyramids when using our method (right column). In contrast, notice the unilateral or bilateral absence of the caudal part of the corticospinal tract (white arrows) when using SFM (middle column) or DTI method (right column).}
\label{fig:tractography_2}
\end{figure*}

\begin{figure*}[htb]
  \centering
\includegraphics[width=1.0\textwidth]{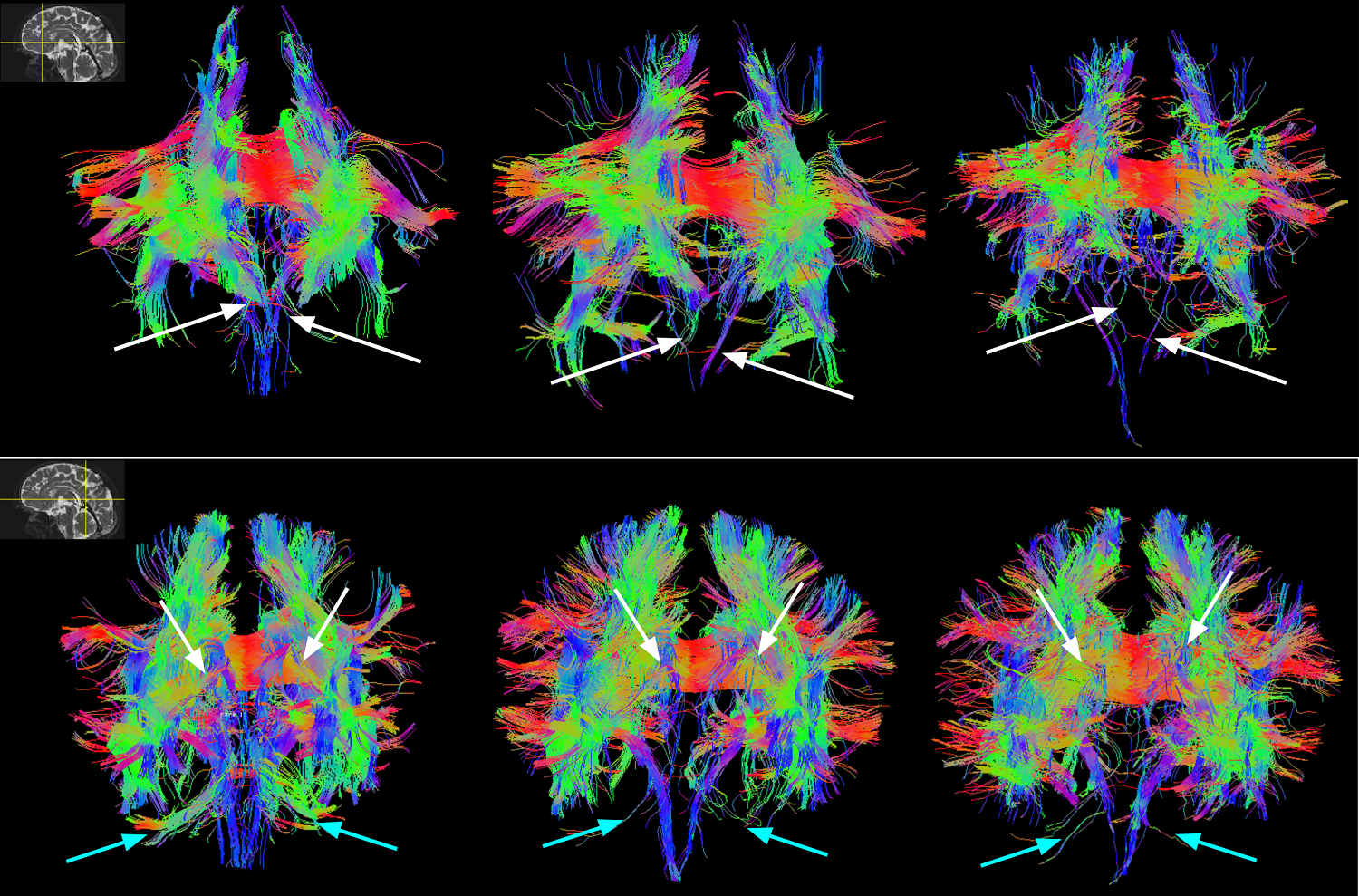}
\caption{Visualization and grading of white matter tracts for our method (left column), SFM (middle column), and DTI (right column) using the frontal (vertical yellow line in the left upper corner of the upper row) and parietal slice filters (vertical line in the left upper corner of the bottom row). Upper row: Note the integrity and bilateral presence of the frontopontine fibers (white arrows) when using our method (left column). In contrast, notice unilateral or absence of the frontopontine fibers (white arrows) when using SFM (middle column) or DTI method (right column). Bottom row: Note the integrity and bilateral presence of all the components of the cingulum (in particular beneath the isthmus of the cingulate gyrus, white arrows) when using our method (left column). In contrast, notice the unilateral absence of components of cingulum (white arrows) when using SFM (middle column) or DTI method (right column). Similarly, notice the bilateral presence of middle cerebellar pedunculi when using our method (left column), and unilateral or (middle column) bilateral absence (right column) when using SFM or DTI.}
\label{fig:tractography_3}
\end{figure*}

We applied our method, SFM, and DTI on 20 subjects from the dHCP dataset and compared the three methods in terms of the quality of the whole-brain tractogram. Overall, as assessed by a neuroanatomist (LV), our method produced more accurate tracts than SFM and DTI. Figures \ref{fig:tractography_1}, \ref{fig:tractography_2}, and \ref{fig:tractography_3} show examples of tracts on which our method resulted in better tract reconstructions than both SFM and DTI.

As we explained in Section \ref{tractography_eval_description}, our expert evaluation of tractograms involved assigning a grade of 1, 2, or 3 to each of 12 separate tracts. For each subject, we summed the scores received by each method on the 12 tracts to arrive at a single overall score in the range $[12, 36]$ for the entire tractogram. On 9 of the 20 subjects, our method obtained a higher score than SFM and DTI. On 5 and 2 of the subjects, respectively, SFM and DTI obtained higher scores. On 3 and 1 of the subjects our method tied as best score with, respectively, SFM and DTI. For a statistical analysis, we applied the Wilcoxon signed-rank test. At a $p$ value of 0.01, the score obtained by our method was significantly higher than both SFM and DTI, and the score of SFM was also significantly higher than DTI.

\section{Conclusions}

In this work we proposed a novel machine learning method for estimating the number and orientations of fascicles in each voxel from DW-MRI measurements. Our method uses all diffusion measurements in a voxel to estimate the angle to the closest fascicle for every direction in a set of directions on the unit sphere. Importantly, this is done separately for each direction. This distinguishes our method from all previous methods that estimate all of their unknown parameters or the distribution function at the same time. For example, for a tensor fitting approach, even with a simplified tensor model as in Equation \eqref{eq:signal_model}, a three-tensor model will have  16 unknown parameters. Tensor-fitting methods solve a non-convex optimization problem to estimate these parameters, which can produce sub-optimal solutions and be sensitive to initialization. This is the main advantage of our method, leading to its superior performance. Our method is much more accurate than several competing methods in detecting voxels with two or three fascicles, which are highly important for tractography and connectivity analysis. Our method is also accurate in estimating fiber orientations, which are also important in tractography. In the absence of ground truth for the number and orientations of fascicles for real data, we showed that our method was more robust than CSD to the reduction of the number of measurements. This makes our method a better choice for applications such as fetal imaging where it is difficult to acquire a large number of high-quality measurements due to fetal movements as well as total acquisition time constraints. In experiments with real data, detailed expert evaluations showed that our method was able to accurately reconstruct various brain tracts. Overall, our experimental results show that the method proposed in this study can be used for accurate estimation of the number and orientations of major fascicles and, hence, for accurate tractography.

\ifCLASSOPTIONcaptionsoff
  \newpage
\fi

\bibliographystyle{IEEEtran}

\bibliography{davoodreferences}

\end{document}